\documentclass[twocolumn,aps,notitlepage,showpacs,showkeys]{revtex4}
\usepackage{amsfonts}
\usepackage{amsmath}
\usepackage{amssymb}
\usepackage{graphicx}
\usepackage{subfigure}
\begin{document}
\title{Universal Quantum Computation in a Neutral Atom Decoherence Free Subspace}
\author{E. Brion}
\email{ebrion@phys.au.dk}
\author{L. H. Pedersen}
\email{lhp@phys.au.dk}
\author{K. M\o lmer}
\affiliation{Lundbeck Foundation Theoretical Center for Quantum
System Research \\ Department of Physics and Astronomy, University
of Aarhus, Ny Munkegade, Bld. 1520, DK-8000 \AA rhus C, Denmark}
\author{S. Chutia and M. Saffman}
\affiliation{Department of Physics, University of Wisconsin, 1150
University Avenue, Madison, Wisconsin 53706 } \keywords{neutral atom
quantum computing, decoherence free subspace} \pacs{03.67.Lx,
32.80.Qk, 32.80.Rm}
\date{\today}

\begin{abstract}
In this paper, we propose a way to achieve protected universal
computation in a neutral atom quantum computer subject to collective
dephasing. Our proposal relies on the existence of a Decoherence
Free Subspace (DFS), resulting from symmetry properties of the
errors. After briefly describing the physical system and the error
model considered, we show how to encode information into the DFS and
build a complete set of safe universal gates. Finally, we provide
numerical simulations for the fidelity of the different gates in the
presence of time-dependent phase errors and discuss their
performance and practical feasibility.

\end{abstract}
\maketitle

\section{Introduction}

Within the last few years, quantum information has become one of the
most promising and active fields in physics. In the commonly used
model, a \emph{quantum computer} consists of two-level systems, the
\emph{qubits}, in which information is stored in a binary fashion.
Calculations on this information are achieved through the
application of particular evolutions of the system, called
\emph{quantum gates} \cite{NC00}. Thanks to \emph{quantum
parallelism} \cite{Deu85},\ which merely follows from the linearity
of quantum mechanics, quantum computers are expected to be much more
efficient than their classical analogues, in particular for
simulating the behaviour of quantum systems \cite{Fey82} and for
solving some "difficult" problems, such as factoring \cite{Sho97}.
Unfortunately, none of the various experimental proposals,
comprising NMR \cite{GC97}, Cavity Quantum Electrodynamics
\cite{DRBH95}, and trapped ion \cite{CZ95} implementations, has
succeeded in fulfilling all the requirements one has to check in
order to design a valuable quantum computer, known as DiVincenzo's
criteria \cite{DiV00}. In particular, \emph{decoherence} which
arises from the interaction of the system with its environment,
remains a major obstacle to the feasibility of quantum computing.\\
\indent Quantum gates may be implemented with neutral atoms using
either short range collisions or long range dipole-dipole
interactions. The proposal of Jaksch \emph{et al.} \cite{JCZRCL00}
suggested using the strong dipole-dipole interactions of highly
excited Rydberg atoms for fast quantum gates. In a recent article
\cite{SW05}, one of the authors provided a detailed study of the
Rydberg scheme using optically trapped $^{87}$Rb atoms. In
particular, it was shown that the errors due to the trap setup
itself can be made quite small by a proper choice of physical
parameters, allowing for fast and reliable single- and two-qubit
gates.\\
\indent In this paper, we elaborate on this proposal by considering
the effect of collective random dephasing errors, which for instance
stem from the uncontrolled action of exterior fields. In Sec. \ref{SysMod}, we briefly
present the physical implementation proposed in \cite{SW05}, as
well as the error model we choose to address and we
show how to protect information through encoding into a Decoherence
Free Subspace (\emph{DFS}), the existence of which merely follows
from the symmetry properties of the errors. In Sec.
\ref{Protection}, we build a
complete set of universal protected gates, which allows us to
perform any computation without leaving the \emph{DFS}. In Sec. \ref{Simulation}, we provide numerical
simulations for the fidelity of the different gates in the presence
of time-dependent phase errors. The practical feasibility of these
gates is then discussed; in particular, the limitation arising from
spontaneous emission is addressed. Finally, in Sec.
\ref{Conclusion}, we give our conclusions and the perspectives of
our work.

\section{A Neutral atom quantum computer \label{SysMod}}

\begin{figure}
[ptb]
\begin{center}
\includegraphics[
height=2.0003in,
width=3in
]%
{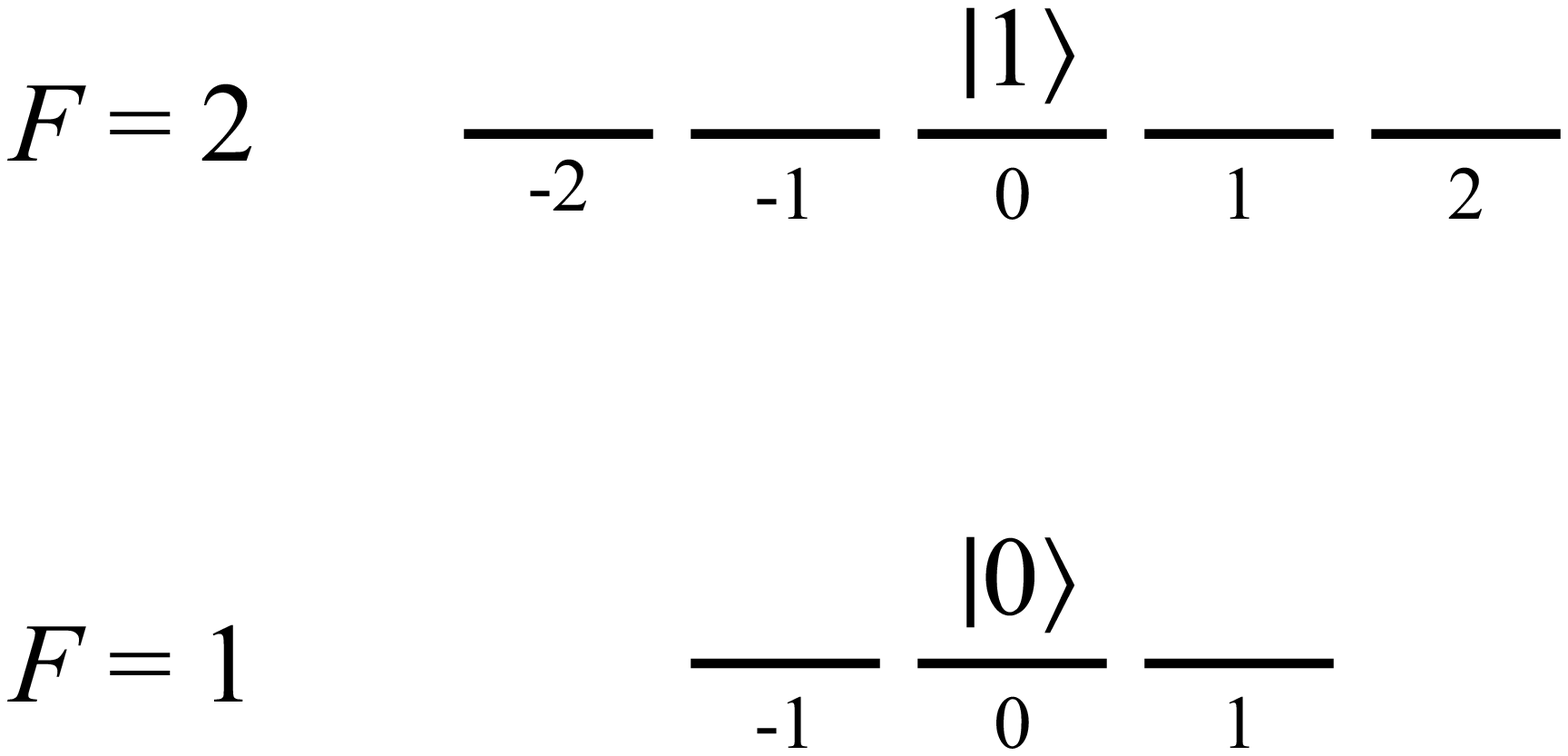}%
\caption{{\small The qubit basis states.}}
\label{qubits}%
\end{center}
\end{figure}

The quantum computing proposal we shall consider throughout this
paper has been put forward recently in \cite{SW05}. The physical
qubit consists of a
$^{87}$Rb atom restricted to the hyperfine states%
\begin{align*}
\left\vert 0\right\rangle  &  \equiv\left\vert 5S_{1/2},F=1,m_{F}%
=0\right\rangle \\
\left\vert 1\right\rangle  &  \equiv\left\vert 5S_{1/2},F=2,m_{F}%
=0\right\rangle
\end{align*}
which form the logical basis (see Fig. \ref{qubits}). After
precooling in a magneto-optical trap (MOT), the different qubits
which constitute the computer are captured in an egg box-style
potential created by a Far-Off-Resonance Trap (FORT) (for a
detailed presentation of the technical aspects of the physical
apparatus, see \cite{SW05}).\\
%EndExpansion
\indent In this setting, the single qubit rotations are implemented
through Raman-like transitions between $\left\vert 0\right\rangle $
and $\left\vert 1\right\rangle $ via an off-resonance excited state,
whereas the two-qubit phase gate is performed according to Jaksch
\textit{et al.}'s proposal \cite{JCZRCL00} which relies on the large
dipole-dipole interaction between Rydberg atoms. Combining
single-qubit rotations and two-qubit phase gates leads to the
standard set of universal quantum gates, comprising the single-qubit
Hadamard $\left(  H\right)  $ and $\pi/8$ $\left( T\right)  $\
gates, as well as the two-qubit $CNOT$ gate :
\begin{equation*}
H\equiv\frac{1}{\sqrt{2}}\left[
\begin{array}
[c]{ll}%
1 & 1\\
1 & -1
\end{array}
\right]  ,\text{ \ }T\equiv\left[
\begin{array}
[c]{ll}%
1 & 0\\
0 & e^{i\pi/4}%
\end{array}
\right]  \] \[ \text{ \ }CNOT\equiv\left[
\begin{array}
[c]{llll}%
1 & 0 & 0 & 0\\
0 & 1 & 0 & 0\\
0 & 0 & 0 & 1\\
0 & 0 & 1 & 0
\end{array}
\right]  .
\end{equation*}
\indent Analysing the error sources due to the trap setup
(background gas collisions, scattering of the trapping light,
heating due to laser noise, ...), the authors of \cite{SW05} show
that, given a proper choice of the physical parameters, single- and
two-qubit gates can be performed at MHz rates with decoherence
probability and fidelity errors at the level of $10^{-3}$.\\
\indent In this paper, we shall consider the effect of random phase
errors which affect different qubits according to the same
(\textit{a priori} time-dependent) error Hamiltonian
\begin{equation*}
\widehat{E}\left(  t\right)  =\hbar\left[
\begin{array}
[c]{ll}%
\epsilon_{0}\left(  t\right)  & 0\\
0 & \epsilon_{1}\left(  t\right)
\end{array}
\right]
\end{equation*}
Typically, this kind of error model stands for the action of
parasitic external fields which induce uncontrolled and unwanted
energy shifts. If no protection scheme is used, the unknown
differential phase shift induced by $\widehat{E}\left( t\right)  $
on the qubit states rapidly leads to a complete loss of coherence. \
\

\section{Protection against the errors through encoding into a \emph{DFS}
\label{Protection}}

\subsection{Secure storage of the information in a \emph{DFS}}

To protect information from quantum errors, one can resort to active
schemes such as quantum codes \cite{NC00}, the most famous example
being the stabilizer codes \cite{Got96}. It is also possible to take
advantage of the symmetry properties of the interaction between the
system and its environment in order to passively protect information
from the effects of decoherence; this is the basic principle of the 
Decoherence Free Subspace ($\emph{DFS}$) strategy \cite{ZR97}, which
consists in encoding information into subspaces immune to errors.
The explicit construction of $\emph{DFS}$ has already been achieved
for certain collective error processes \cite{DG97}, and even
experimentally implemented in quantum optics \cite{KBA00} and
trapped ion \cite{KMR01} setups. Moreover, universal computation
within these $\emph{DFS}$'s has been shown to be possible from a
theoretical point of view \cite{VK03}.\ Here, we shall deal with the
practical implementation of such a \emph{DFS} in
the neutral atom system described in \cite{SW05}.\\
\indent In the case of collective dephasing, a $\emph{DFS}$ can be
straightforwardly identified in the Hilbert space of a
two-physical-qubit system. Indeed, as the two qubit states $\left\{
\left\vert 0_{\emph{DFS}}\right\rangle \equiv \left\vert
01\right\rangle ,\left\vert 1_{\emph{DFS}}\right\rangle
\equiv\left\vert 10\right\rangle \right\}  $ are affected in the
same way by phase errors
\begin{align*}
\left\vert 0_{\emph{DFS}}\right\rangle  &  \equiv\left\vert 01\right\rangle
\overset{t}{\rightarrow}e^{-i\int_{0}^{t}\left(  \epsilon_{0}\left(
\tau\right)  +\epsilon_{1}\left(  \tau\right)  \right)  d\tau}\left\vert
01\right\rangle \\
\left\vert 1_{\emph{DFS}}\right\rangle  &  \equiv\left\vert 10\right\rangle
\overset{t}{\rightarrow}e^{-i\int_{0}^{t}\left(  \epsilon_{0}\left(
\tau\right)  +\epsilon_{1}\left(  \tau\right)  \right)  d\tau}\left\vert
10\right\rangle
\end{align*}
the subspace they span is clearly left invariant by the errors. This
shows that a qubit of information can be safely stored in the state
of a two-physical-qubit system.\\
\indent From a practical point of view, in order to protect the
information contained in a physical qubit initially in the state
$\left\vert \psi\right\rangle =c_{0}\left\vert 0\right\rangle
+c_{1}\left\vert 1\right\rangle $, one simply adds a second
auxiliary qubit, initially prepared in the state $\left\vert
1\right\rangle $ and performs a $CNOT$ gate, which achieves the
$\emph{DFS}$
encoding%
\[
\left\vert \psi\right\rangle \left\vert 1\right\rangle =c_{0}\left\vert
01\right\rangle +c_{1}\left\vert 11\right\rangle \overset{CNOT}{\rightarrow
}c_{0}\left\vert 01\right\rangle +c_{1}\left\vert 10\right\rangle =\left\vert
\psi_{\emph{DFS}}\right\rangle .
\]
Decoding is straightforwardly performed through applying the $CNOT$
gate again. Moreover, a protected $N$-logical-qubit memory can be
constructed just by associating $N$ protected cells, which
requires $2N$ physical qubits.\\
\indent To process the information safely stored in the $\emph{DFS}$
we now need to design a new set of universal gates, comprising the
single-logical-qubit Hadamard and $\pi/8$ gates, as well as the
two-logical-qubit $CNOT$ gate. This point is actually not obvious.
Indeed, the gate implementations proposed in \cite{SW05} cannot be
used any longer as they would make the system leave the
$\emph{DFS}$, resulting in erroneous calculations. We thus have to
resort to new physical primitives. In the following we review two
secure processes which constantly remain in the $\emph{DFS}$. We
then show how to combine them to implement universal protected
computation.

\subsection{Physical primitives for universal protected computation}

\begin{figure}
[ptb]
\begin{center}
\includegraphics[
height=2.0003in,
width=3in
]%
{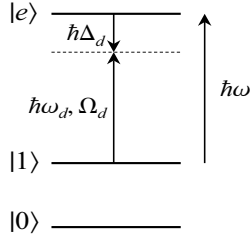}%
\caption{{\small Level scheme for the dephasing gate.}}
\label{figdeph}%
\end{center}
\end{figure}

\begin{figure}\centering
%[ptb]
\includegraphics[width = 10cm]%
{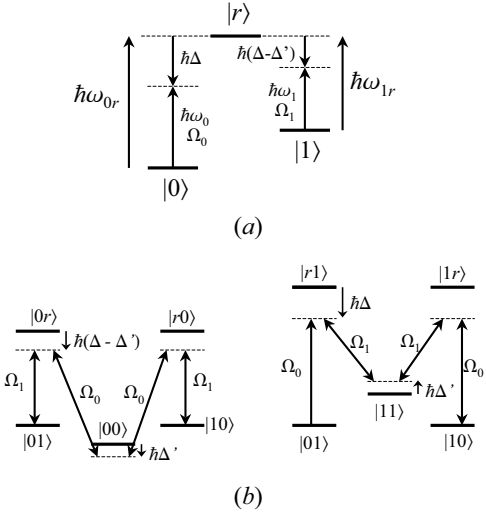}%
\caption{{\small a) Level scheme and laser couplings for one atom.
b) The two `$M$' paths coupling $\left\vert 01\right\rangle $ and
$\left\vert 10\right\rangle $.\ }}
\label{4photongate}%
\end{figure}

The first primitive consists of a phase gate, obtained through
applying a laser to the first atom of the logical qubit. The Rabi
frequency is denoted $\Omega_{d}$ and the laser frequency
$\omega_{d}$\ is detuned by the quantity $\Delta_{d}$ from the
frequency $\omega$\ of the transition $\left\vert 1\right\rangle
\rightarrow\left\vert e\right\rangle $ where $\left\vert
e\right\rangle $ denotes an excited state
($\Delta_{d}\equiv\omega_{d}-\omega$, see Fig. \ref{figdeph}). After
a time $t_d = \frac{n2\pi}{\Omega_{R,d}}$, where
$\Omega_{R,d}=\sqrt{|\Omega_d|^2+\Delta_d^2}$ and $n$ is a positive
integer, the state $|10\rangle$ picks up the phase factor
$e^{i\varphi}$, where $\varphi=n\pi\left(
1+\frac{\Delta_d}{\Omega_{R,d}}\right)$, while the state $\left\vert
01\right\rangle $ is left unchanged. In the \emph{DFS} basis, this
transformation is represented by the matrix \
\[
P\left(  \varphi\right)  =\left[
\begin{array}
[c]{ll}%
1 & 0\\
0 & e^{i\varphi}%
\end{array}
\right]
\]%
%TCIMACRO{\FRAME{ftbpFU}{3in}{2.0003in}{0pt}{\Qcb{Dephasing gate : level
%scheme.}}{\Qlb{figdeph}}{figdeph.eps}{\special{ language "Scientific Word";
%type "GRAPHIC";  maintain-aspect-ratio TRUE;  display "USEDEF";
%valid_file "F";  width 3in;  height 2.0003in;  depth 0pt;
%original-width 0pt;  original-height 0pt;  cropleft "0";  croptop "1";
%cropright "1";  cropbottom "0";
%filename 'FigDeph.eps';file-properties "XNPEU";}}}%
%BeginExpansion
%EndExpansion
Given an arbitrary phase $\varphi_0$ one can choose the different
physical parameters such that $\varphi = \varphi_0$ and $|\Delta_d|
\gg |\Omega_d|$, which ensures that the excited state $|e\rangle$ remains
essentially unpopulated during the process.\\
\indent The second primitive is implemented
by applying two laser fields $\overrightarrow{E}_{i}\left(
t\right)  =\frac{1}{2}\left(
\overrightarrow{\underline{E}}_{0,i}e^{-i\left(  \omega_{i}t-\varphi
_{i}\right)  }+cc\right)  $, $i=1,2$, to a pair of atoms. The
frequencies $\omega_{i}$ are assumed slightly detuned from the
transitions $\left\vert i\right\rangle \rightarrow\left\vert
r\right\rangle $, $i=0,1$, respectively, where $\left\vert
r\right\rangle $ is a Rydberg state of the atom. The
detunings are denoted (see Fig. \ref{4photongate}a))%
\[
\Delta = \omega_0-\omega_{0r} \quad \text{ and\ } \quad \Delta' =
\omega_0-\omega_1-\omega_{01}
\]
where $\omega_{ir}$ is the frequency of the transition $\left\vert
i\right\rangle \rightarrow\left\vert r\right\rangle $. Let us
emphasize that the two atoms involved in the transformation are not
bound to belong to the same logical qubit, \textit{i.e.} they do not
have to be in a superposition of the states $\left\vert
01\right\rangle $ and $\left\vert 10\right\rangle $.\\
\indent As is well known \cite{Gallagher}, Rydberg atoms exhibit
huge dipole moments which lead to large dipole-dipole interactions
responsible for dipole blockade \cite{JCZRCL00}. We shall assume
that the dipole-dipole interaction is only significant when both
atoms are in their Rydberg states $\left\vert
r\right\rangle $. In other words, we shall assume $\widehat{V}_{dd}%
=\hbar\Delta_{rr}\left\vert rr\right\rangle \left\langle
rr\right\vert $. When the fields are applied to a pair of atoms,
transitions may occur between the different resonant physical qubit
product states, in agreement with energy conservation. We observe
that the states $\left\vert 00\right\rangle $ and $\left\vert
11\right\rangle $ only couple to themselves to second order, and, as
the doubly excited Rydberg state $\left\vert rr\right\rangle $ is
shifted far off resonance due to the dipole-dipole interaction, the
only two paths actually coupling $\left\vert 01\right\rangle $ and
$\left\vert 10\right\rangle $ are the two fourth-order `$M$' systems
drawn on Fig. \ref{4photongate}b).\\ 
\indent Assuming $ \left \vert \Omega_{0} \right \vert,\left \vert \Omega_{1} \right \vert \ll \left \vert \Delta \right \vert, \left \vert  \Delta^{\prime} \right \vert, \left \vert \Delta^{\prime} - \Delta \right \vert ,  \left \vert  \Delta_{rr} \right \vert$, one can extract the effective dynamics of the subspace $\left\{  \left\vert
00\right\rangle ,\left\vert 01\right\rangle ,\left\vert
10\right\rangle ,\left\vert 11\right\rangle \right\}  $ through a perturbative approach \cite{Faisal}
which yields the evolution operator, expressed in the interaction picture,  
\begin{equation*}
U_{eff}\left(  t,\Omega_{0},\Omega_{1},\Delta,\Delta^{\prime},\Delta
_{rr}\right)  = e^{-i\left(  \Delta_{0}-\Delta^{\prime}\right) t}
\] \[ \times
\left[
\begin{array}
[c]{llll}
e^{-i\left(  \Delta_{00}-\Delta_{0}\right)  t} & 0 & 0 & 0\\
0 & \cos\left(  \frac{\Omega_{R}t}{2}\right)   & i\sin\left(  \frac{\Omega
_{R}t}{2}\right)   & 0\\
0 & i\sin\left(  \frac{\Omega_{R}t}{2}\right)   & \cos\left(  \frac{\Omega
_{R}t}{2}\right)   & 0\\
0 & 0 & 0 & e^{-i\left(  \Delta_{11}-\Delta_{0}\right)  t}%
\end{array}
\right]
\end{equation*}
where
\begin{widetext}
\begin{eqnarray}
\Omega _{R} &=&\frac{\left\vert \Omega _{0}\Omega _{1}\right\vert ^{2}}{8}%
\frac{\Delta _{rr}\left( \Delta ^{\prime }-2\Delta \right) }{\left( \Delta
_{rr}+\Delta ^{\prime }-2\Delta \right) \Delta ^{2}\left( \Delta ^{\prime
}-\Delta \right) ^{2}} \label{OmR}\\
\Delta _{0} &=&\frac{\left\vert \Omega _{0}\right\vert ^{2}}{4\Delta }+\frac{%
\left\vert \Omega _{1}\right\vert ^{2}}{4\left( \Delta -\Delta ^{\prime
}\right) }-\frac{1}{16}\left[ \frac{\left\vert \Omega _{0}\right\vert ^{4}}{%
\Delta ^{3}}+\frac{\left\vert \Omega _{1}\right\vert ^{4}}{\left( \Delta
-\Delta ^{\prime }\right) ^{3}}+\frac{\left\vert \Omega _{0}\Omega
_{1}\right\vert ^{2}\left( 2\Delta _{rr}+\Delta ^{\prime }-2\Delta \right)
\left( 2\Delta -\Delta ^{\prime }\right) }{\left( \Delta _{rr}+\Delta
^{\prime }-2\Delta \right) \Delta ^{2}\left( \Delta ^{\prime }-\Delta
\right) ^{2}}\right] \label{D0} \\
\Delta _{00} &=&\frac{\left\vert \Omega _{0}\right\vert ^{2}}{2\Delta }\left[
1+\frac{1}{2\Delta }\left( \frac{\left\vert \Omega _{1}\right\vert ^{2}}{%
2\Delta ^{\prime }}-\frac{\left\vert \Omega _{0}\right\vert ^{2}\left(
\Delta _{rr}-\Delta \right) }{\Delta \left( \Delta _{rr}-2\Delta \right) }%
\right) \right]  \label{D00} \\
\Delta _{11} &=&\frac{\left\vert \Omega _{1}\right\vert ^{2}}{2\left( \Delta
-\Delta ^{\prime }\right) }\left[ 1+\frac{1}{2\left( \Delta ^{\prime
}-\Delta \right) }\left( \frac{\left\vert \Omega _{0}\right\vert ^{2}}{%
2\Delta ^{\prime }}+\frac{\left\vert \Omega _{1}\right\vert ^{2}\left(
\Delta _{rr}+\Delta ^{\prime }-\Delta \right) }{\left( \Delta -\Delta
^{\prime }\right) \left( \Delta _{rr}+2\Delta ^{\prime }-2\Delta \right) }%
\right) \right] \label{D11}
\end{eqnarray}
\end{widetext}
\indent It is important to note that if the two atoms involved are
initially prepared in a $\emph{DFS}$ state, \textit{i.e.} in a
superposition of $\left\vert 01\right\rangle $ and $\left\vert
10\right\rangle $, the transformation $U_{eff}$ leaves them in the
$\emph{DFS}$ and the associated gate
\[
R\left(  \theta\right)  \propto\left[
\begin{array}
[c]{ll}%
\cos\left(  \theta\right)  & i\sin\left(  \theta\right) \\
i\sin\left(  \theta\right)  & \cos\left(  \theta\right)
\end{array}
\right]  ,\text{ with \ }\theta\equiv\frac{\Omega_{R}t}{2}.
\]
is thus safe. But it is also interesting to note that if the two
atoms considered are not in a $\emph{DFS}$\ state, as for example in
the state $\left\vert 00\right\rangle $, they pick up a phase
factor, which can be controlled via the Rabi frequencies and
detunings. This observation will be used in the following to
implement a
two-logical-qubit phase gate.\\
%EndExpansion
\indent To conclude, let us emphasize that the transformations we
have just considered are not completely \emph{error-free}. They are
\emph{safe} in the sense that they do not make the system leave the
\emph{DFS}, the information thus being constantly protected from the
effect of the errors. But the errors will of course affect the
primitives $P$ and $U_{eff}$ (and thus the gates we shall
build by combining them) through the detunings $\Delta,\Delta^{\prime}%
,\Delta_{rr}$ which will take slightly different values from the
ideally expected ones. However, the detunings being large, the
effect of reasonably small errors on the system will be quite
unnoticeable. The limitation of the fidelity for the quantum gates
will be numerically analyzed in the following section.

\subsection{Construction of the universal safe quantum gates}
Let us now see how the previous primitives can be used to implement
universal safe computation. The single-logical-qubit $\pi/8$ $\left(
T\right)  $\ and Hadamard $\left(H\right)$ gates are
straightforwardly obtained according to
\begin{equation*}
T_{DFS}\equiv\left[
\begin{array}
[c]{ll}%
1 & 0\\
0 & e^{i\frac{\pi}{4}}%
\end{array}
\right]  \propto P\left(  \frac{\pi}{4}\right)  \]
\[\text{ \
}H_{DFS}\equiv\frac {1}{\sqrt{2}}\left[
\begin{array}
[c]{ll}%
1 & 1\\
1 & -1
\end{array}
\right]  \propto P\left(  \frac{3\pi}{2}\right)  \cdot R\left(  \frac{\pi}%
{4}\right)  \cdot P\left(  \frac{3\pi}{2}\right)
\end{equation*}
\indent We now consider two logical qubits, consisting of the
physical qubits $\left( 1,2\right)  $ and $\left(  3,4\right)  $
respectively. The two-logical-qubit $\emph{DFS}$\ basis is
\begin{align*}
\left\vert 00\right\rangle _{\emph{DFS}}  &  \equiv\left\vert
0101\right\rangle \\
\left\vert 01\right\rangle _{\emph{DFS}}  &  \equiv\left\vert
0110\right\rangle \\
\left\vert 10\right\rangle _{\emph{DFS}}  &  \equiv\left\vert
1001\right\rangle \\
\left\vert 11\right\rangle _{\emph{DFS}}  &  \equiv\left\vert
1010\right\rangle .
\end{align*}

\begin{figure}
 \includegraphics[width=1.8cm]{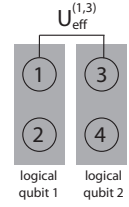}
 \caption{\small The $U_{eff}$ gate acting on atoms 1 and 3.}
 \label{qubitsu}
\end{figure}
One might think that in order to implement a logical two-qubit gate
a four-atom process must be involved. It turns out, however, that a
two-atom operation is sufficient to implement a controlled phase
gate between two logical qubits. Thus consider the transformation
$U_{eff}^{\left(  1,3\right)  }\left(  \tau,\Omega
_{0},\Omega_{1},\Delta,\Delta^{\prime},\Delta_{rr}\right)  $
involving atoms $\left(  1\right)  $ and $\left(  3\right)  $ from
different $\emph{DFS}$ pairs, cf. Fig. \ref{qubitsu}. This operation
leaves the subspace $\left\{  \left\vert 00\right\rangle
_{\emph{DFS} },\left\vert 01\right\rangle _{\emph{DFS}},\left\vert
10\right\rangle _{\emph{DFS}},\left\vert 11\right\rangle
_{\emph{DFS}}\right\}  $ invariant and is given by

\[
U_{eff}^{\left(  1,3\right)  }\left(  t,\Omega_{0},\Omega_{1},\Delta
,\Delta^{\prime},\Delta_{rr}\right)  =e^{-i\left(
\Delta_{0}-\Delta^{\prime }\right)  t}\] \[\times \left[
\begin{array}
[c]{llll}%
e^{-i\left(  \Delta_{00}-\Delta_{0}\right)  t} & 0 & 0 & 0\\
0 & \cos\left(  \frac{\Omega_{R}t}{2}\right)   & i\sin\left(  \frac{\Omega
_{R}t}{2}\right)   & 0\\
0 & i\sin\left(  \frac{\Omega_{R}t}{2}\right)   & \cos\left(  \frac{\Omega
_{R}t}{2}\right)   & 0\\
0 & 0 & 0 & e^{-i\left(  \Delta_{11}-\Delta_{0}\right)  t}%
\end{array}
\right]
\]
in the $\emph{DFS}$ basis. If we adjust the different parameters in
such a way that
\begin{equation}
t=\tau=\frac{4\pi}{\Omega_{R}},\text{\ }\frac{\Delta_{00}-\Delta_{0}}%
{\Omega_{R}}=\frac{k}{2},\text{\ }\frac{\Delta_{11}-\Delta_{0}}{\Omega_{R}
}=\frac{1}{4}+\frac{l}{2} \label{ConditionPhase}
\end{equation}
where $k$, $l$ are integers, we get\ \
\[
U_{eff}^{\left(  1,3\right)  }\left(  \tau,\Omega_{0},\Omega_{1},\Delta
,\Delta^{\prime},\Delta_{rr}\right)  \propto\left[
\begin{array}
[c]{llll}%
1 & 0 & 0 & 0\\
0 & 1 & 0 & 0\\
0 & 0 & 1 & 0\\
0 & 0 & 0 & -1
\end{array}
\right]  \equiv U_{phase,\emph{DFS}}%
\]
corresponding to the conditional phase gate in the $\emph{DFS}$ subspace. It
is now straightforward to implement the protected $CNOT$ gate
\[
CNOT_{\emph{DFS}}=\left(  I\otimes H_{\emph{DFS}}\right)  \cdot
U_{phase,\emph{DFS}}\cdot\left(  I\otimes H_{\emph{DFS}}\right)
\]
\begin{figure}
 \includegraphics[width=7.1cm]{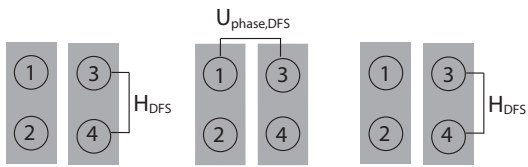}
 \caption{\small The $CNOT_{DFS}$ gate.}
 \label{qubitscnot}
\end{figure}
cf. Fig. \ref{qubitscnot}, which completes our set of protected
universal quantum gates.

\section{Numerical simulations \label{Simulation}}
As already pointed out, the gates we have just implemented are not
error-free, as their fidelities are limited by the errors which
modify the detunings involved. Spontaneous emission from the excited states $\left \vert e \right \rangle$ and $\left \vert r \right \rangle$ also restricts the performances of our gates, and has to be taken into account. In this section, we numerically
estimate the influence on the fidelity of random phase errors and spontaneous emission and
compare the protected gates to a set of unprotected gates. Finally
we discuss the practical interest and feasibility of our proposal.

\subsection{Performances of the protected gates}
\begin{figure} \centering
\includegraphics[width=7cm]{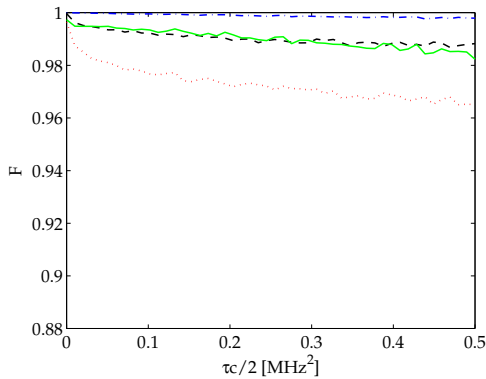}
\caption{{\small Performance of the universal set of gates without
spontaneous emission. Dash-dotted: $T_{DFS}$, dashed: $H_{DFS}$,
solid line: $U_{phase,DFS}$, dotted: $CNOT_{DFS}$. Parameters for
the gates: $P$ gate: $\Omega_d/2\pi = 5 $ MHz, $\Delta_d/2\pi
\approx -49.81 $ MHz, $t(\pi/4) \approx 1.00$ $\mu$s.
$R\left(\frac{\pi}{4}\right)$ gate: $\Omega_0/2\pi \approx 3.91 $
MHz, $\Omega_1/2\pi \approx 1.97 $ MHz, $\Delta/2\pi \approx 60.34 $
MHz, $\Delta'/2\pi \approx 30.70 $ MHz, $\Delta_{rr}/2\pi = 100 $
MHz, $t \approx 120.20 $ $\mu$s. $U_{phase,DFS}$ gate:
$\Omega_0/2\pi \approx 3.93 $ MHz, $\Omega_1/2\pi \approx 1.96 $
MHz, $\Delta/2\pi \approx 60.48 $ MHz, $\Delta'/2\pi \approx 30.05 $
MHz, $\Delta_{rr}/2\pi = 100 $ MHz, $t \approx 938.62 $ $\mu$s. For
the generation of the phase errors the parameters are: $\tau = 1$
$\mu$s, $\alpha_e = \alpha_r = 1.5$.}}\label{gates}
\end{figure}

\begin{figure} \centering
\includegraphics[width=7cm]{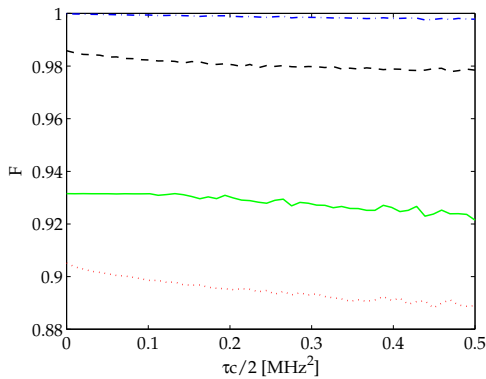}
\caption{{\small Performance of the gates including
spontaneous emission with $\gamma_r/2\pi = 5 $ kHz. Parameters and notations are the same as for
Fig. \ref{gates}. }}\label{gatess}
\end{figure}

In this subsection, we first describe how phase errors and spontaneous emission are modelled in our system. Then we introduce the fidelity measure we shall use throughout this section. Finally we explain how simulations are performed, and discuss our results in detail.\\
\indent Here we assume phase errors on $|1\rangle$, $|e\rangle$ and
$|r\rangle$ and model their effect by the single-atom Hamiltonian
\begin{equation*} \widehat{E}\left(  t\right)
=\hbar\left[ \epsilon_1 (t) |1\rangle \langle 1| + \epsilon_e(t)
|e\rangle \langle e| + \epsilon_r(t) |r\rangle \langle r| \right]
\end{equation*}
Moreover, we suppose that phase errors can be described by an
Ornstein-Uhlenbeck process for which \cite{Gil95}:
\begin{equation*}
\epsilon(t+dt) =
\epsilon(t)-\frac{1}{\tau}\epsilon(t)dt+\sqrt{c}G(t)\sqrt{dt}
\end{equation*}
where $\tau$ is the relaxation time, $c$ is the diffusion constant
and $G(t)$ is the unit Gaussian variable. The Ornstein-Uhlenbeck process is characterized
by the correlation function 
\[
\left\langle \epsilon\left( t \right)  \epsilon\left( t' \right)  \right\rangle = \frac{\tau c}{2} e^{- \left\vert t-t' \right\vert / \tau}
\]
in steady state. We assume that the errors for the various levels
are correlated, which seems reasonable for background magnetic field 
perturbations. Thus $\epsilon_1(t)$ is
generated using the approach described above while $\epsilon_e(t)$
and $\epsilon_r(t)$ are found from $\epsilon_e(t) = \alpha_e
\epsilon_1(t)$ and $\epsilon_r(t) = \alpha_r \epsilon_1(t)$.\\
\indent Spontaneous emission from the excited state (used for the
implementation of the $P$ gate) and the Rydberg state can be taken
into account by adding the term $-i\hbar \left(
\frac{\gamma_e}{2}|e\rangle \langle e|+ \frac{\gamma_r}{2}|r\rangle
\langle r|\right)$ to the single-atom Hamiltonian. For the numerical
implementation we have considered $|e\rangle = |r\rangle$. The
radiative line width of the Rydberg state, $\gamma_r/2\pi$, is
estimated to be smaller than 10 kHz for $n > 65$ \cite{SW05}.\\
\indent An appropriate fidelity measure for the performance of the
gates in the presence of stochastic errors is calculated in the
following way. For an $N\times N$ density matrix, $\rho$, the time
evolution is given by $\rho \xrightarrow{U} U\rho U^\dagger$.
Representing the matrix $\rho$ as a vector $\overrightarrow{\rho}$
this can be rewritten such that $\overrightarrow{\rho}
\xrightarrow{U} M \overrightarrow{\rho}$ with $M$ being an
$N^2\times N^2$ matrix. To each instance of
$\epsilon_1(t)$ (and correspondingly $\epsilon_e(t)$ and
$\epsilon_r(t)$) corresponds a certain $M$. In order to obtain an
average over the various phase error configurations, we can now
simply average over the corresponding $M$s and subsequently
calculate the fidelity from:
\begin{equation*}
F \left( \left\vert \psi \right\rangle \right)  =  \overrightarrow{\rho}^\dagger_{ideal}
M_{av} \overrightarrow{\rho}
\end{equation*}
where $\rho_{ideal} = U^{\phantom{\dagger}}_{ideal}\rho
U^\dagger_{ideal}$, $M_{av}$ denotes the average over $M$ and $\rho=|\psi\rangle \langle \psi|$. In practice, we evaluate $F$ for a large number of pure states $\left\vert \psi \right\rangle$, and in the figures we present the minimum value of $F$ 
over these input states.\\
\indent As shown in the previous section, the different universal gates can be obtained by combination of the two primitives $P\left(\varphi\right)$ and $U_{eff} \left ( t, \Omega_{0}, \Omega_{1}, \Delta, \Delta^{\prime}, \Delta_{rr} \right )$: it is thus necessary to perform these primitives with high precision. The expression obtained for $P\left( \varphi \right)$ is exact: it implies that, in absence of errors and spontaneous emission, if we tune $\left\{ n, \Omega_{d}, \Delta_{d} \right \}$ such that $\varphi = n \pi \left( 1 + \frac{\Delta_{d}}{\sqrt{\left \vert \Omega_{d} \right \vert ^{2} + \left \vert \Delta_{d} \right \vert ^{2} }} \right)$ is the desired phase factor, we shall exactly get the expected phase gate $P\left( \varphi \right)$. On the contrary, the expression we obtained for $U_{eff}$ is only perturbative. It means that if we choose the physical parameters $\left\{ t, \Omega_{0}, \Omega_{1}, \Delta, \Delta^{\prime}, \Delta_{rr}  \right\}$ so that the effective parameters Eq.(\ref{OmR}-\ref{D11}) take the desired values (to implement a conditional phase gate or an $R\left( \theta \right)$ gate), the exponential of the full-Hamiltonian will be slightly different from the desired gate, even in the ideal case. Before dealing with errors and spontaneous emission, we refine the parameters by a numerical search in the neighborhood of our first guess provided by the analytical perturbative approach. Once we have a faithful gate in the ideal case, we keep the same set of parameters and include the new terms discussed above in the Hamiltonian in order to run our simulations.\\
\indent Fig. \ref{gates} and \ref{gatess} show the performances of
the different gates when spontaneous emission is not/is included,
respectively. The fidelity is plotted as a function of $\frac{\tau
c}{2}$, which corresponds to the variance of the detuning due to perturbations.
As is evident in Fig. \ref{gatess} spontaneous emission severely limits the fidelity.\\
\indent Note that in \cite{SW05} the transitions to the Rydberg
states are implemented by going off-resonantly via
$|5P_{1/2}\rangle$. This worsens the problem of spontaneous
emission, as the $R$ and $U_{phase,DFS}$ gates are thus quite slow
and the radiative linewidth for the $|5P_{1/2}\rangle$ state is
quite large ($\gamma_{5P_{1/2}}/2\pi=5.7$ MHz). It is thus
necessary to be strongly detuned from the level $|5P_{1/2}\rangle$,
but as this reduces the coupling to the Rydberg levels, the
intensity of the Raman beams should at the same time be
increased.\\
\indent A further source of decoherence for the experimental setup
proposed in \cite{SW05} is the motion of atoms in the traps: as
the interatomic distance, $R$, varies so does the dipole-dipole
interaction, which scales as $1/R^3$. Based on estimates provided in
\cite{SW05} a variation of 20 \% for the value of $\Delta_{rr}$ is
expected. We performed a numerical simulation assuming that
$\Delta_{rr}$ is harmonically varying which demonstrated a dramatic
reduction of the fidelity for realistic parameters. Atomic motion in
the traps thus poses a serious limitation for the gates. To improve
the performances of the gates the atoms might be cooled further or
the distance between the traps could be increased, which would,
however, reduce $\Delta_{rr}$.

\subsection{Comparison with unprotected gates}
\begin{figure}
\includegraphics[width=7cm]{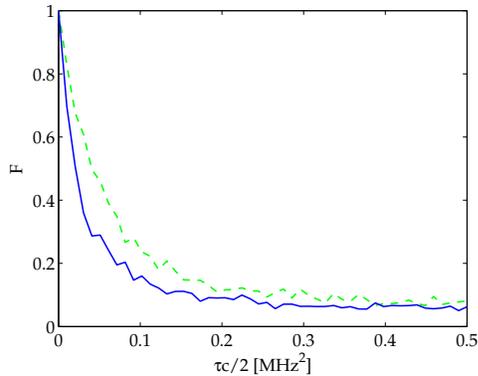}
\caption{{\small Performance for the unprotected gates. Dashed:
$H_{DFS}$, solid line: $CNOT_{DFS}$. Parameters: $\Omega/2\pi = 0.5$
MHz, $\Delta_{rr}/2\pi = 100$ MHz, $\tau = 1$
$\mu$s, $\alpha_e = \alpha_r = 1.5$.}}\label{upgates}
\end{figure}

In order to assess the performance of the universal set of {\it DFS}
gates suggested in this paper, we compare them to a universal set of
unprotected gates.\\
\indent If we remove the restriction that the system should remain
in the {\it DFS} at all times, a controlled-phase gate between two
atoms can be carried out as suggested in \cite{JCZRCL00} in the
regime of a large interaction strength. Combining this gate with two
single-atom Hadamard gates, which are obtained through single-atom
rotations between states $|0\rangle$ and $|1\rangle$ via an excited
state, results in a two-atom $CNOT$ gate. A universal set of gates
is then given by:
\begin{align*}
H_{DFS} &= CNOT_{1,2}\cdot (H\otimes I)\cdot CNOT_{1,2}\\
CNOT_{DFS} &= CNOT_{1,3}\cdot CNOT_{1,4}
\end{align*}
where $CNOT_{i,j}$ is a $CNOT$ gate with atom $i$ being the control
qubit and atom $j$ being the target qubit. The $T_{DFS}$ gate is
simply implemented in the same way as for the protected gates.\\
\indent In Fig. \ref{upgates} we plot the performance of the
unprotected Hadamard and $CNOT$ gates. As expected, these gates are
much more sensitive to phase errors than the protected gates.
Spontaneous emission from the Rydberg states also severely limits
the fidelity for the unprotected gates, as the controlled-phase gate
involves resonant transitions to the Rydberg levels.\\
\indent The durations of the gates are, however, smaller than for
the protected gates. Thus, with the parameters for Fig.
\ref{upgates} the durations of the gates are $t(H_{DFS}) =
15.5$ $\mu$s and $t(CNOT_{DFS})
= 14$ $\mu$s.\\
\indent Another advantage of the unprotected gates as opposed to the
protected ones we have suggested is that they are less sensitive to
variations in $\Delta_{rr}$, and thus to atomic motion.

\section{Conclusion \label{Conclusion}}

In this paper we have identified a set of logical qubit basis states
for a neutral atom decoherence free subspace and a corresponding set
of protected universal gates. Numerical simulations demonstrate that
the proposed set of gates is much more robust against phase errors
than a set of unprotected gates. They are, however, also much more
affected by the motion of the atoms in the traps. Therefore, one
has to consider what the worst source of decoherence is for a given
physical situation and subsequently choose which set of gates to
use. Even if the set of protected gates is assessed to be
unfavorable, one should still consider using the logical qubit bases
states for computations, as information is then protected during
storage of the
qubits.\\
\indent The work presented in this paper naturally leads to further
developments. First, we would like to study resource optimised
$\emph{DFS}$ encoding. Here, in order to build a reliable
$N=2$-qubit register\ we have merely associated two
two-physical-qubit protected cells, but it is clear that for an
$N$-qubit quantum memory $\left( N\geq3\right)$ the number of
physical qubits required by this straightforward scheme,
\textit{i.e.} $2N$ physical qubits, is much larger than actually
needed. This new construction of a $\emph{DFS}$ will induce new
problems, as regards practical encoding and processing of the
information stored in the $\emph{DFS}$, the resolution of which will
probably require the use of quantum control techniques, such as
nonholonomic control \cite{NHC}. Moreover, we also plan to extend
our results to more general error models as for instance
position-dependent errors.

\section{Acknowledgements}
This work was supported by ARO-DTO grant nr. 47949PHQC. S.C. and M.S. were also supported by NSF grant PHY-0205236 and K.M. was supported by the European Union integrated project SCALA.

\end{document}